\pdfoutput=1
\documentclass[11pt, a4paper]{article}
\usepackage{relsize}
\usepackage{array}
\usepackage{a4wide}
\usepackage[hidelinks]{hyperref}
\usepackage{latexsym}
\usepackage{cite}
\usepackage{mathrsfs}
\usepackage{amssymb}
\usepackage{amsmath}
\usepackage{amsfonts}
\usepackage{amsthm}
\usepackage{graphicx}
\usepackage{enumerate}
\usepackage[dvipsnames]{xcolor}
\usepackage{todonotes}
\usepackage{bm}
\usepackage{cancel}
\usepackage{float}
\usepackage{verbatim}
\usepackage{mathtools}
\usepackage{subcaption}
\usepackage{xcolor}
\usepackage{tikz}
\usepackage[compat=1.1.0]{tikz-feynman}
\graphicspath{ {Images/} }

\tikzfeynmanset{warn luatex=false}
\renewcommand{\vec}{\bs}  
\newcommand{\bs}[1]{\boldsymbol{#1}}

\usepackage{etoolbox}
\apptocmd{\sloppy}{\hbadness 10000\relax}{}{}

\newcommand{\bea}{\begin{eqnarray}\label}
\newcommand{\eea}{\end{eqnarray}}

\makeatletter
\newcommand*{\textoverline}[1]{$\overline{\hbox{#1}}\m@th$}
\newcommand*\bigcdot{\mathpalette\bigcdot@{.65}}
\newcommand*\bigcdot@[2]{\mathbin{\vcenter{\hbox{\scalebox{#2}{$\m@th#1\bullet$}}}}}
\makeatother

\definecolor{darkgreen}{RGB}{0,120,0}

\newcommand{\eq}[1]{\begin{equation}#1\end{equation}}
\newcommand{\eqs}[1]{\begin{equation}\begin{split}#1\end{split}\end{equation}}

\date{}

\begin{document}

\title{Soft Photon, Gluon and Graviton Theorems in (A)dS from Conformal Invariance}

\author{Jiajie Mei${}^{a}$ and Yuyu Mo${}^b$ \vspace{7pt}\\ \normalsize  \textit{
${}^a$ Institute of Physics}\\\normalsize\textit{University of Amsterdam, Amsterdam, 1098 XH, The Netherlands}\\ 
\textit{
${}^b$
Higgs Centre for Theoretical Physics}\\\normalsize\textit{School of Physics and Astronomy},\\\normalsize\textit{
The University of Edinburgh, Edinburgh EH9 3FD, Scotland, UK}}

{\let\newpage\relax\maketitle}
\begin{abstract}
We present new soft theorems for photon, gluon, and graviton correlators at tree level in (Anti)-de Sitter space. The results are derived by applying conformal Ward identities to constrain the structure of Witten diagrams.
\end{abstract}

\author{}

\pagebreak
\tableofcontents
\section{Introduction}
Soft graviton theorem at the leading order was first derived by Weinberg \cite{Weinberg:1965nx}. This beautiful and universal result has shown to be highly effective in constraining the S-matrix \cite{Hodges:2012ym} and is understood to encode the asymptotic symmetries of flat spacetime\cite{He:2014laa,Strominger:2013jfa}, specifically the Bondi-Metzner-Sachs (BMS) symmetry\cite{Bondi:1962px,Sachs:1962wk}. Beyond leading order, soft theorems in flat space can be derived from Poincaré symmetry and gauge invariance; in particular, \cite{Bern:2014vva,Broedel:2014fsa} showed that on-shell gauge Ward identities—building on earlier ideas from \cite{Low:1958sn,Gross:1968in}—can fully determine the (sub)-subleading soft theorems. Furthermore, this approach leads to an infinite set of soft theorems, as demonstrated in \cite{Li:2018gnc,Hamada:2018vrw}, although they only constrain part of the full amplitude.\\
From the cosmology side, the infinite set of Ward identities was understood even earlier \cite{Creminelli:2012ed,Hinterbichler:2013dpa,Hinterbichler:2012nm} and originates from the concept of adiabatic modes introduced by Weinberg \cite{Weinberg:2003sw}. However, these consistency conditions appear to differ from the soft theorems in flat space beyond leading order\cite{Mirbabayi:2016xvc}. \\
More recently, a connection was established in (A)dS between shift symmetries in the bulk and enhanced soft limits of boundary correlators \cite{Armstrong:2022vgl}, which generalizes the known relationship between soft theorems and symmetries in flat-space S-matrix. This suggests that soft limits of boundary correlators should also have a symmetry-based interpretation. Our goal in this paper is to derive the soft theorems for spinning correlators using amplitude-based methods, see also \cite{McFadden:2014nta,Chowdhury:2024wwe,Chowdhury:2024snc,Albayrak:2024ddg}.

For the S-matrix, we perfer to make Lorentz symmetry manifest, and then the only remaining constraint is the gauge Ward identity. Solving this identity in the soft expansion yields the sub-leading soft theorems. We use a similar logic to derive soft theorems in (A)dS. However, there are key differences: first, after constructing a general diagrammatic ansatz for the boundary correlator and fixing the gauge, it is the isometry group of (A)dS that constrains the Witten diagrams, rather than gauge Ward identities. Second, unlike in flat space, the propagators in (A)dS are not divergent in the soft limit. As a result, the leading soft theorem in (A)dS appears at order $q^0$ where $q$ is the soft momentum, and all diagrams contribute to the leading soft behavior—unlike in flat space, where only diagrams with soft propagators dominate and yield a $q^{-1}$ divergence.

In this paper, by solving the special conformal Ward identity in the soft expansion, we derive the leading soft theorems for photons and gluons. Beyond leading order, the soft behavior is only partially constrained—similar to the situation in flat-space amplitudes—and we explicitly write down the resulting infinite partial soft theorems. For Einstein gravity, both the leading and subleading soft theorems are derived and fully determined.

\section{Conformal Ward Identities}
Our convention for the metric on $\mathrm{AdS}_{d+1}$ is:
\begin{equation}
ds^2=\frac{\mathcal{R}^2}{z^{2}} (dz^{2}+\eta_{\mu \nu} dx^\mu dx^\nu) ,
\end{equation}
with $0 < z < \infty$ and $\mathrm{AdS}$ radius $\mathcal{R}=1$. The boundary operators obey conformal symmetry; in particular, the special conformal generators in Fourier space are given by,
\eq{
K_{\mu} \mathcal{O}_{\Delta}^{(\ell)}=\left(2(\Delta-d) \partial_{k_\mu}+k^\mu\partial_{k^\rho} \partial_{k_\rho}-2 k^\rho\partial_{k^\rho} \partial_{k_\mu}-2 \epsilon_\rho \partial_{k^\rho} \partial_{\epsilon_\mu}+2 \epsilon_\mu \partial_{k^\rho} \partial_{\epsilon_\rho}\right) O_{\Delta}^{(\ell)}.
}
This implies that the boundary correlators in $\mathrm{AdS}$ (or similarly, the wavefunction coefficients in $\mathrm{dS}$~\cite{Baumann:2020dch,Baumann:2022jpr,Arkani-Hamed:2018kmz}) obey conformal Ward identities,
\eqs{
\sum_{a=1}^n K_a^\mu  \langle \mathcal{O}_1 \dots \mathcal{O}_a \dots \mathcal{O}_n \rangle =0.
}
However, unlike in the S-matrix case, the right-hand side of the Ward identities for correlators is often non-vanishing, typically involving terms supported on delta functions. Upon Fourier transforming, these delta functions translate into local terms in momentum space. This poses a significant obstacle when attempting to interpret or uncover symmetries directly from correlators.\\
For example, the relation between enhanced soft limits and symmetries~\cite{Armstrong:2022vgl} holds only upon imposing the free equations of motion on the bulk-to-boundary propagators and discarding total derivative terms via integration by parts, thereby neglecting all local terms.

To overcome this issue, we employ the idea of the generalized LSZ reduction formula, which effectively amputates the external states by applying the equations of motion. Hence, in our analysis, we focus exclusively on the on-shell part of the correlators~\cite{Cheung:2022pdk}. This approach is particularly transparent and effective within the Mellin-Momentum formalism~\cite{Mei:2023jkb,Mei:2024abu,Mei:2024sqz}:
\begin{align}\label{eq:MMamplitude}
   \langle \mathcal{O}_1 \dots \mathcal{O}_n \rangle & =  \int [ds_i]\int \frac{dz}{z^{d+1}} \mathcal{A}_n(z\vec k,s)\prod_{i=1}^n\phi(s_i,k_i)z^{-2 s_i+d/2},
\end{align}
with  
$
\int [ds_i] = \prod_{i=1}^n \int_{-i \infty}^{+i \infty} \frac{d s_i}{2 \pi i}.
$
We perform the Mellin transform of the standard bulk-to-boundary propagator \cite{Sleight:2019mgd,Sleight:2019hfp},
$
\phi_{\Delta}(k, z) = \int_{-i \infty}^{+i \infty} \frac{d s}{2 \pi i}\, z^{-2 s + d/2} \, \phi_{\Delta}(s, k).
$
Once $\mathcal{A}_n$ is determined, the correlator can be straightforwardly obtained by following the prescription outlined in~\cite{Mei:2024sqz}. Thus, $\mathcal{A}_n$ will be our primary focus, which we refer to as the on-shell amplitude. We can subsequently define a new operator that acts solely on $\mathcal{A}_n$ by commuting through the external states:
\eqs{
 (\mathcal{K}^{\mu} \mathcal{A}_n) \phi  := K^{\mu} (\mathcal{A}_n \phi).
}
We find that
\eqs{
	\mathcal{K}^\mu \mathcal{A} &
 ={\color{black}\frac{\left(\left(d-2 \Delta\right)^2-16 s^2\right)}{4 k^2}k^\mu\mathcal{A} +\left[2\left(2s-d /2\right)\partial_{k_\mu} +k^\mu\partial_{k^\rho} \partial_{k_\rho}-2 k^\rho\partial_{k^\rho} \partial_{k_\mu}+\Sigma^{\mu}\right]\mathcal{A}},\\
& \Sigma^{\mu}=-2\epsilon_{\rho}\left( \frac{k^{\rho}(\Delta-d/2-2s)}{k^2}+\partial_{k^{\rho}}\right) \partial_{\epsilon_{\mu}}+2\epsilon^{\mu}\left( \frac{k^{\rho}(\Delta-d/2-2s)}{k^2}+\partial_{k^{\rho}}\right) \partial_{\epsilon_{\rho}}.
\label{scwispinpart}
}
The conformal Ward identity should now be exactly zero, as shown in \cite{Mei:2024sqz},
\eqs{
\sum_{a=1}^n \mathcal{K}_a^{\mu} \mathcal{A}_n=0.
}
\subsection{Helicity basis}
Spinning correlators can always be split into two parts, the transverse-traceless part and the longitudinal part \cite{Bzowski:2013sza}.  In our analysis, we aim to constrain only the transverse-traceless part, as the longitudinal component can always be reconstructed from the gauge Ward identity for the conserved current. In practice, we work in the helicity basis, defined as:
\eqs{
\varepsilon_{\mu}(k)&\equiv \epsilon^{\nu}\Pi_{\mu \nu}(k)=\epsilon^{\nu} \left(\eta_{\mu \nu}-\frac{k_\mu k_\nu}{k^2} \right),\\
\varepsilon_{\mu \nu}(k) \equiv \epsilon^{\rho \sigma}\Pi_{\mu \nu, \rho \sigma}(k)&=\epsilon^{\rho \sigma} \left(  \frac{1}{2} \Pi_{\mu \rho}\Pi_{\nu \sigma}+ \frac{1}{2}\Pi_{\mu \sigma} \Pi_{\rho \nu}-\frac{1}{d-1}  \Pi_{\mu \nu}\Pi_{\rho \sigma} \right).
\label{projectors}
}
So $\epsilon_i \cdot k_i \neq 0$ but $\varepsilon_i \cdot k_i =0$. Restricting to spin-1 and spin-2 conserved currents (with $\Delta = d + J - 2$), we can then rewrite the generator in \eqref{scwispinpart} on only tranverse-traceless part as:
\eqs{
\mathcal{K}^\mu\left(\epsilon^\sigma\Pi_{\sigma \rho}(k)\mathcal{A}^\rho(k)\right )&= \mathcal{K}^{\mu}_J (\varepsilon_{\rho}\mathcal{A}^\rho(k)),\\
\mathcal{K}^\mu(\epsilon^{\lambda}\epsilon^\nu\Pi_{{\lambda} \nu, \rho \sigma}(k)\mathcal{A}^{\rho\sigma}(k))&= \mathcal{K}_J^{\mu} (\varepsilon_{\rho}\varepsilon_\sigma\mathcal{A}^{\rho\sigma}(k)).
}
We then obtain the generator \(\mathcal{K}_J^\mu\), which is more convenient as it ensures that the momentum derivative acts solely on the amplitude, and not on the helicity $\varepsilon$:
\eqs{
 \mathcal{K}^\mu_J \mathcal{A} 
 =&{\color{black} \frac{\left((4-2J-d)^2-16 s^2\right)}{4 k^2}k^\mu\mathcal{A} +\left[2\left(2s-d /2\right)\partial_{k_\mu} +k^\mu\partial_{k^\rho} \partial_{k_\rho}-2 k^\rho\partial_{k^\rho} \partial_{k_\mu}+\Sigma^{\mu}\right]\mathcal{A}},\\
&\Sigma^{\mu}=-2\varepsilon_{\rho}\partial_{k^{\rho}}\partial_{\varepsilon_{\mu}}+2\varepsilon^{\mu}\left(\frac{k^{\rho}(J-2-d/2-2s)}{k^2}+\partial_{k^{\rho}} \right) \partial_{\varepsilon_{\rho}}, 
\label{eq10}}
where \(J\) is the spin, taking values 1 or 2.

\section{Soft Photon theorem}
In this section, we focus on photons interacting with \(\Delta = d-1\) scalars, with one of the photons being soft. As shown in Figure \ref{fig:class_I_and_Class_2}, there are two types of diagrams that contribute to this amplitude:
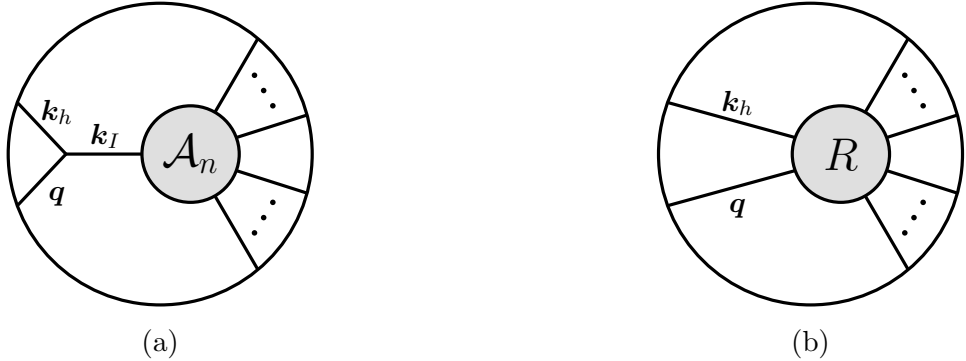
\begin{figure}[htbp]
	\centering
	\begin{minipage}{0.45\textwidth}
		\centering
		\begin{tikzpicture}[baseline,scale=0.8]
			\draw[very thick, fill=lightgray!50] (0.5, 0) circle (0.8);
			\draw[very thick] (0, 0) circle (2.5);
			
			\draw[very thick, black] (-1.55, 0) -- (-0.3,0);
			\draw[very thick, black] (-1.55, 0) -- ({-2.5*cos(20)}, {2.5*sin(20)});
			\draw[very thick, black
            ]
			({-2.5*cos(-20)}, {2.5*sin(-20)})-- 	(-1.55, 0) ;
			
			
			\coordinate (Big1) at ($(0,0)+(2.5*cos{50},2.5*sin{50})$);
			\coordinate (Small1) at ($(0.5,0)+(0.8*cos{60},0.8*sin{60})$);
			\draw[very thick,  black
            ]
			(Small1) -- (Big1);
			\coordinate (Big2) at ($(0,0)+(2.5*cos{15},2.5*sin{15})$);
			\coordinate (Small2) at ($(0.5,0)+(0.8*cos{20},0.8*sin{20})$);
			\draw[very thick,  black
            ]
			(Small2) -- (Big2);
			
			\coordinate (Bigdd) at ($(0.5,0)+(1.6*cos{20},1.6*sin{20})$);
			\coordinate (Smalldd) at ($(0.5,0)+(1.8*cos{60},1.8*sin{60})$);
			\foreach \t in {0.25, 0.5, 0.75}
			{\fill ($(Bigdd)!\t!(Smalldd)$) circle (1.3pt);}
			\coordinate (Bigdd) at ($(0.5,0)+(1.6*cos{20},-1.6*sin{20})$);
			\coordinate (Smalldd) at ($(0.5,0)+(1.8*cos{60},-1.8*sin{60})$);
			\foreach \t in {0.25, 0.5, 0.75}
			{\fill ($(Bigdd)!\t!(Smalldd)$) circle (1.3pt);}
			
			\coordinate (Big11) at ($(0,0)+(2.5*cos{-50},2.5*sin{-50})$);
			\coordinate (Small11) at ($(0.5,0)+(0.8*cos{-60},0.8*sin{-60})$);
			\draw[very thick]
			(Small11) -- (Big11);
			\coordinate (Big22) at ($(0,0)+(2.5*cos{-15},2.5*sin{-15})$);
			\coordinate (Small22) at ($(0.5,0)+(0.8*cos{-20},0.8*sin{-20})$);
			\draw[very thick]
			(Small22) -- (Big22);
			
			\node at (0.5, 0) {\LARGE $\mathcal{A}_{n}$};
			\node at (-1.7, 0.7) {$\vec k_h$};
			\node at (-1.7, -0.7) {$\vec q$};
			\node at (-0.9, 0.3) {$\vec k_I$};
		\end{tikzpicture}\vspace{0.7em}
		\par (a)
	\end{minipage}
	\hfill
	\begin{minipage}{0.45\textwidth}
		\centering
		\begin{tikzpicture}[baseline,scale=0.8]
		\draw[very thick, fill=lightgray!50] (0.5, 0) circle (0.8);
		\draw[very thick] (0, 0) circle (2.5);
				\coordinate (Bigguabf) at ($(0,0)+(2.5*cos{-160},2.5*sin{-160})$);
		\coordinate (Smallguabf) at ($(0.5,0)+(0.8*cos{-160},0.8*sin{-160})$);
			\draw[very thick,  black
            ] (Bigguabf) -- (Smallguabf);
			\coordinate (Bigscalar) at ($(0,0)+(2.5*cos{160},2.5*sin{160})$);
			\coordinate (Smallscalar) at ($(0.5,0)+(0.8*cos{160},0.8*sin{160})$);
			\draw[very thick] (Bigscalar) -- (Smallscalar);  
	\coordinate (Big1) at ($(0,0)+(2.5*cos{50},2.5*sin{50})$);
	\coordinate (Small1) at ($(0.5,0)+(0.8*cos{60},0.8*sin{60})$);
	\draw[very thick,  black
    ]
	(Small1) -- (Big1);
	\coordinate (Big2) at ($(0,0)+(2.5*cos{15},2.5*sin{15})$);
	\coordinate (Small2) at ($(0.5,0)+(0.8*cos{20},0.8*sin{20})$);
	\draw[very thick,  black
    ]
	(Small2) -- (Big2);
	
	\coordinate (Bigdd) at ($(0.5,0)+(1.6*cos{20},1.6*sin{20})$);
	\coordinate (Smalldd) at ($(0.5,0)+(1.8*cos{60},1.8*sin{60})$);
	\foreach \t in {0.25, 0.5, 0.75}
	{\fill ($(Bigdd)!\t!(Smalldd)$) circle (1.3pt);}
	\coordinate (Bigdd) at ($(0.5,0)+(1.6*cos{20},-1.6*sin{20})$);
	\coordinate (Smalldd) at ($(0.5,0)+(1.8*cos{60},-1.8*sin{60})$);
	\foreach \t in {0.25, 0.5, 0.75}
	{\fill ($(Bigdd)!\t!(Smalldd)$) circle (1.3pt);}
	
	\coordinate (Big11) at ($(0,0)+(2.5*cos{-50},2.5*sin{-50})$);
	\coordinate (Small11) at ($(0.5,0)+(0.8*cos{-60},0.8*sin{-60})$);
	\draw[very thick]
	(Small11) -- (Big11);
	\coordinate (Big22) at ($(0,0)+(2.5*cos{-15},2.5*sin{-15})$);
	\coordinate (Small22) at ($(0.5,0)+(0.8*cos{-20},0.8*sin{-20})$);
	\draw[very thick]
	(Small22) -- (Big22);
			
			\node at (0.5, 0) {\LARGE $R$};
			\node at (-1.2, 0.85) {$\vec k_h$};
			\node at (-1.2, -0.9) {$\vec q$};
		\end{tikzpicture}\vspace{0.7em}
		\par (b)
	\end{minipage}
	\caption{(a): Diagram with three-point interaction and bulk-to-bulk propagator with $\vec{k}_I=\vec{k}_h +\vec{q}$ (b): other type of diagram denoted by $R$}
    \label{fig:class_I_and_Class_2}
\end{figure}

\eqs{
\mathcal{A}_{n+1}(\vec{q},\vec{k}_1,\dots,\vec{k}_n)=&\sum_h e_h z \varepsilon_q \cdot k_h G(\vec{k}_I,z,z') \mathcal{A}_n(z',\vec k_1,\ldots,\vec{k}_I, \dots,\vec{k}_n) 
\\
&\hspace{1cm}+ R (\vec{q},\vec{k}_1,\dots, \vec{k}_h, \dots, \vec{k}_n),
\label{eq11}
}
where $\vec k_I\equiv \vec q+\vec k_h$ and we make use of the three-point photon-scalar interaction \( e_h z \varepsilon_q \cdot k_h \), with \( e_h \) being the coupling constant. The hard particle label $h$ runs over all hard scalar legs.
 It is straightforward to verify that all other conformal symmetries are manifest and that there is no gauge redundancy in the helicity basis. We are then left with the constraint from the special conformal Ward identity:
\eqs{
\left( \sum_{a=1}^{n} \mathcal{K}_a^{\mu} +\mathcal{K}_q^{\mu} \right )\mathcal{A}_{n+1}=0.
\label{eq13constrain}
}
As a second-order differential equation, this is notoriously difficult to solve. However, we will focus solely on the $\varepsilon_q^{\mu}$ component of the vector constraints. 
\\
For diagram (a) in Figure \ref{fig:class_I_and_Class_2}, only part of the following two operators will contribute to $\varepsilon_q^{\mu}$:
\eqs{
&(\mathcal{K}^{\mu}_q +\mathcal{K}^{\mu}_h)\left( e_h z \varepsilon_s \cdot k_h G(\vec{k}_I,z,z')  \mathcal{A}_n(z',\vec{k}_1, \dots,\vec{k}_I, \dots, \vec{k}_n)  \right)\\
=&-\varepsilon_q^\mu \frac{4 s_q+d-2}{2 zq^2}\Big(z^2 k_I^2+2 (s_q+s_h) \big(d-2 (s_q+s_h+1)\big)\Big)  G(\vec{k}_I,z,z')  e_h \mathcal{A}_n\\
=& \,\varepsilon_q^{\mu} e_h\left(-\frac{4s_q+d-2}{2z q^2} \mathcal{A}_n(z,\vec{k}_1, \dots,\vec{k}_I, \dots, \vec{k}_n)  \right).
\label{eq_ideneity_op_photon}
}
In the second step, we use integration by parts, after which the EoM operator cancels the bulk-to-bulk propagator by definition,
\eqs{
&\mathcal{D}_{k_I}^{\Delta} G(\vec{k}_I,z,z')=\delta(z-z'),\\
&\mathcal{D}_{k_I}^{\Delta}   \equiv   z^{2}k_I^{2}-z^{2}\partial_{z}^{2}-(1-d)z\partial_{z}+\Delta(\Delta-d).
\label{Ddef}
}
More details are explained in Appendix \ref{appendix:A}. As for diagram (b) in Figure \ref{fig:class_I_and_Class_2}, focusing on the coefficients of polarization basis $\varepsilon_q^\mu$ only, we have,
\eqs{
\mathcal{K}^{\mu}_q (\varepsilon_q \cdot R)=&\,\varepsilon_q^{\mu} \left( -\frac{4s_q+d-2}{q^2} q \cdot R +2  \partial_q \cdot R    \right).
\label{classIIscwi}
}
How other $\mathcal{K}_a^\mu$ acts on the $R$ function will not be important as it has no $\frac{1}{q^2}$ pole, which in turn contributes at different orders in the soft expansion.
Finally, our main constraint equation is,
\eqs{
-\frac{4s_q+d-2}{2z q^2} \Big (\sum_h e_h \mathcal{A}_n(z,\vec{k}_1, \dots ,\vec{q}+\vec{k}_h, \dots ,\vec{k}_n) +2z q \cdot R\Big) +2 \partial_q \cdot R .
   \label{contrainseq}
}
\subsection{Solving SCWI in soft expansion}
Our goal is to solve the constraint equation \eqref{contrainseq} recursively in the soft expansion. To this end, we consider the following Taylor expansion:
\eqs{
 \mathcal{A}_n(z,\vec{k}_
 1, \dots, \vec{q}+\vec{k}_h, \dots, \vec{k}_n)  & = \sum_{\ell =0}^{\infty} \frac{1}{\ell !} (q \cdot  \partial_{k_h})^{\ell} \mathcal{A}_n(z,\vec{k}_
 1, \dots, \vec{k}_h, \dots, \vec{k}_{n}) \\
  R(z,\vec{q},\vec{k}_1, \dots, \vec{k}_n)  &= \varepsilon_q \cdot R^{(0)} (z,\vec{k}_1, \dots, \vec{k}_n)+ \varepsilon_q^{\mu} q^{\nu} R^{(1)}_{\mu \nu}(z,\vec{k}_1, \dots, \vec{k}_n) + \dots\\
  &\hspace{1cm}+ \varepsilon_q^{\mu} q^{\nu_1} \dots q^{\nu_\ell} R^{(\ell )}_{ \mu \nu_1 \dots \nu_\ell}  (z,\vec{k}_1, \dots, \vec{k}_n)+\ldots.
  \label{AandRexpansion}
}
Now we are ready to solve the constraints order by order in $\vec{q}$. Due to the $\frac{1}{q^2}$ pole, the leading order is $\mathcal{O}(q^{-2})$ and can only come from eq.~(\ref{eq_ideneity_op_photon}):
\eqs{
-\sum_{h}e_h\frac{\varepsilon_q^{\mu}}{2zq^2}(4s_q+d-2)\mathcal{A}_n=0.
}
This implies charge conservation,
\eqs{
\sum_{h}e_h =0.
}
At order $\mathcal{O}(q^{-1})$,  both diagrams in Figure \ref{fig:class_I_and_Class_2} contribute,
\eqs{
-\frac{\varepsilon_q^{\mu}}{q^2}(4s_q+d-2) \left( \sum_h e_h \frac{q \cdot \partial_{k_h} \mathcal{A}_n}{2z} +q \cdot R_0 \right)=0.
}
Solving this constraint leads to:
\eqs{
R_{\mu}^{(0)}=-\sum_h e_h\frac{1}{2z} \partial_{k_h^\mu} \mathcal{A}_n.
}
We can now proceed to solve equation \eqref{contrainseq} to higher orders, but clearly starting from order $\mathcal{O}(q^0)$, many other operators can also contribute. However, the unknown function $R^{(0)}$ is already solved above, so we can recursively solve $R$ by focusing on the terms containing the $1/q^2$ pole:
\eqs{
-\frac{4s_q+d-2}{q^2} \left ( \sum_h e_h \frac{(q\cdot \partial_{k_h})^{\ell+1}}{2z (\ell+1)!} \mathcal{A}_n  + q^{\mu} q^{\nu_1} \dots q^{\nu_\ell} R^{(\ell)}_{ \mu \nu_1 \dots \nu_\ell}  \right).
}
We find,
\eqs{
R^{(\ell)}_{ \mu \nu_1 \dots \nu_\ell} =-\sum_h e_h \frac{1}{2z(\ell+1)!}\partial_{k^{\mu}_h} \partial_{k^{\nu_1}_h} \dots \partial_{k^{\nu_\ell}_h} \mathcal{A}_n + N_{\mu \nu_1 \dots \nu_\ell},
\label{RRRphoton}
}
with $N$ being arbitrary antisymmetric tensor in terms of swapping $\mu \leftrightarrow \nu _i$. Plugging this solution into the special conformal Ward identity, we check that it automatically solves all the $\varepsilon_q^\mu$ part of the constraint. So we find the infinite soft theorem at the integrand level that is given by:
\eqs{
\mathcal{A}_{n+1}=&\sum_{\ell =0 }^\infty\sum_h e_h \left( z' \varepsilon_q \cdot k_h G(\vec{k}_I,z,z') - \frac{1}{2 z(\ell+1)} \varepsilon_q \cdot \partial_{k_h} \right) \frac{1}{\ell !}(q \cdot \partial_{k_h})^{\ell} \mathcal{A}_n(z)\\
&+ \varepsilon_q^{\mu} q^{\nu_1} \dots q^{\nu_{\ell}}N_{\mu \nu_1 \dots \nu_{\ell}}.
\label{mmamp_soft_photon_theorem}
}
Next, we also perform a soft expansion of \(G(\vec{k}_I,z',z)\) and integrate it out \cite{Chowdhury:2024wwe}. However, since only the leading-order amplitude is completely determined here, we present only the leading-order result for correlator here. Restoring the bulk-to-boundary propagator and bulk integral:
\eqs{
 \int_0^{\infty} \frac{d z}{z^{d+1}} z\varepsilon_q \cdot k_h \phi_{d-1}\left(z, k_h\right) \phi_{d-1}\left(z, q\right)  G\left(\vec k_h,z, z^\prime\right)=-\frac{\mathcal{N}_{d-1}}{2} \varepsilon_q \cdot \partial_{k_h} \phi_{d-1}\left(z^\prime , k_h\right),
\label{back_to_momentum_class1_spin1}
}
where $\mathcal{N}_{d-1}=\frac{2^{(d-3) / 2} \Gamma\left(\frac{d-2}{2}\right)}{\sqrt{\pi}}$ is the normalization constant from the bulk-to-boundary propagator. Then we can put the result back into the correlator definition in eq.~(\ref{eq:MMamplitude}), to obtain the final result for correlators in momentum space,
\eqs{
\langle J(\vec q) \phi(\vec k_1) \dots \phi(\vec k_h) \dots J(\vec k_n)\rangle =-\frac{1}{2}\sum_h e_h \varepsilon_q \cdot \partial_{k_h} \langle \phi(\vec k_1) \dots \phi(\vec k_h) \dots J(\vec k_n) \rangle+ \mathcal{O}(q).
\label{eq25}
}
We have verified this formula up to five points.
\section{Soft Gluon theorem}
We consider the color-ordered gluon amplitude, whose on-shell three-point function in the helicity basis is simply:
\eqs{
\mathcal{A}_3(1,2,3)=z(\varepsilon_1 \cdot \varepsilon_2 \varepsilon_3 \cdot k_1+\varepsilon_2 \cdot \varepsilon_3 \varepsilon_1 \cdot k_2 + \varepsilon_3 \cdot \varepsilon_1 \varepsilon_2 \cdot k_3).
\label{3ptym}
}
Then the diagram in Figure~\ref{fig:class_I_and_Class_2}~(a) can be determined by factorization \cite{Mei:2024abu} as follows\footnote{Note that this diagram also includes an additional term, which comes either from taking the OPE limit or directly from the covariant spinning propagator. However, since these terms do not involve a bulk-to-bulk propagator, they can always be absorbed into the remainder function. As we will see later, the solution for the remainder function reproduces exactly such terms. }:
\eqs{
& \sum_{r=\pm} \mathcal{A}_3(\vec q,\vec k_h,-\vec k_I^r) \frac{1}{\mathcal{D}_{\vec k_I}^{d-1}} \mathcal{A}_n(\vec k_I^{-r},\dots, \vec{k}_{n})=\mathcal{A}^{\mu}_3(\vec q,\vec k_h,-\vec k_I) \frac{ \Pi_{\mu \nu}(\vec k_I) }{\mathcal{D}_{\vec k_I}^{d-1}}  \mathcal{A}^{\nu}_n(\vec k_I,\dots, \vec{k}_{n}),
\label{classIym}
}
where the polarization sum $\Pi_{\mu \nu}$ is the projector defined in eq.~\eqref{projectors}, and $\frac{1}{\mathcal{D}_{\vec{k}_I}^{d-1}}$ is defined as the insertion of the scalar bulk-to-bulk propagator,
\begin{equation}
    (\mathcal{D}_{\vec{k}}(z))^{-1} \mathcal{O}(z) = \int \frac{d y}{y^{d+1}} G(\vec{k}, z, y) \mathcal{O}(y).
    \label{DtoG}
\end{equation}
Substituting the 3-point \eqref{3ptym} into \eqref{classIym}, we get,
\eqs{
&z( \varepsilon_q \cdot k_h \varepsilon_h^{\mu}+ \varepsilon_q^{\mu_1}  q^{\mu_2} (\Sigma_{\mu_1 \mu_2})_{ \mu_3 \mu_4} \varepsilon_h^{\mu_3} \eta^{\mu \mu_4} )\Pi_{\mu \nu}(\vec k_I)\mathcal{A}^{\nu}_n(z',\vec k_I,\ldots, \vec{k}_{n}) \frac{1}{\mathcal{D}_{\vec k_I}^{d-1}}
\label{class1_ym}
}
where $\left(\Sigma^{\mu \rho}\right)^{\mu_i \alpha} \equiv \left(\eta^{\mu \mu_i} \eta^{\alpha \rho}-\eta^{\mu \alpha} \eta^{\mu_i \rho}\right)$. With the lesson from photon example, we now focus on the terms with \( \frac{1}{q^2} \) pole in the vector basis \( \varepsilon_q^\mu \) of \( (\mathcal{K}_q^\mu+\mathcal{K}_h^\mu ) \) acting on~\eqref{class1_ym} gives:
\eqs{
-\frac{4s_q+d-2}{2zq^2}\varepsilon_q^{\mu} \left(\varepsilon_h^{\rho} \Pi_{\rho \sigma}(\vec k_I) \mathcal{A}^{\sigma}_n(z',\vec k_I,\ldots, \vec{k}_{n})  \right)
\label{eq:1c_gluon}
}
where we used $\mathcal{D}_k^{d-1}$ to cancel the bulk-to-bulk propagator exactly the same as the photon example. For color ordering, the hard leg label $h$ can only be $1$ or $n$. By summing their contributions, together with the $R$ term, we obtain the constraint equation:
 \eqs{
- \frac{4s_q + d - 2}{2z q^2} \varepsilon_q^{\mu} 
\Big(&
    \varepsilon_h^{\rho} \Pi_{\rho \sigma}(\vec{q} + \vec{k}_1) \mathcal{A}_n^{\sigma}(z, \vec{q} + \vec{k}_1, \ldots, \vec{k}_{n}) 
   \\ &\hspace{1cm} -
    \varepsilon_n^{\rho} \Pi_{\rho \sigma}(\vec{q} + \vec{k}_n) \mathcal{A}_n^{\sigma}(z,  \vec{k}_1, \ldots,\vec{q} +\vec{k}_{n}) 
    + 2z q \cdot R
\Big).
\label{eq:alll_gluon}
}
\subsection{Solving SCWI in soft expansion}
The soft expansion is
\eqs{
\varepsilon_h^{\mu} \Pi_{\mu \nu}(\vec{q} + \vec{k}_h)\mathcal{A}^{\nu}_n(z,\vec{q} + \vec{k}_h,\dots, \vec{k}_{n}) & = \sum_{\ell=0}^{\infty} \frac{\varepsilon_h^{\mu}}{\ell !} (q \cdot \partial_{k_h})^{\ell} \Pi_{\mu \nu}(\vec{k}_h)\mathcal{A}^{\nu}_n(z,\vec{k}_h,\dots, \vec{k}_{n}) \\
&=\sum_{\ell=0}^{\infty} \frac{1}{\ell !} (q \cdot \partial_{k_h})^{\ell}\mathcal{A}_n(z,\vec{k}_h,\dots, \vec{k}_{n}),
\label{a_soft_expansion}
}
where the derivative should also act on the helicity $\varepsilon$. Then eq. (\ref{eq:alll_gluon}) becomes:
\eqs{ 
- \frac{4s_q + d - 2}{2z q^2} \varepsilon_q^{\mu} \Big(&
   \sum_{\ell=0}^{\infty} \frac{1}{\ell!}\left(q \cdot \partial_{k_1}\right)^{\ell} \mathcal{A}_n\left(\vec{k}_1, \ldots, \vec{k}_n\right)
    -
   \sum_{\ell=0}^{\infty} \frac{1}{\ell!}\left(q \cdot \partial_{k_n}\right)^{\ell} \mathcal{A}_n\left(\vec k_1, \ldots, \vec k_n\right)
   \\ &\hspace{1cm} + \sum_{\ell=0}^{\infty} 2z q^\mu q^{\nu_1} \ldots q^{\nu_{\ell}} R_{\mu \nu_1 \ldots \nu_{\ell}}^{(\ell)}\left(z, \vec{k}_1, \ldots, \vec{k}_n\right)
\Big) = 0,
}
The leading order, $\mathcal{O}(q^{-2})$, vanishes automatically. At the next order, $\mathcal{O}(q^{-1})$, the solution reads:
\eqs{
R_{\mu}^{(0)}=&\frac{1}{2z}(\partial_{k_n^{\mu}} -\partial_{k_1^{\mu}} ) \mathcal{A}_n,\\
=&\frac{1}{2z}(\varepsilon_{n\nu}\partial_{k_n^{\mu}}  \mathcal{A}_n^{\nu} - \frac{\varepsilon_n^{\mu}}{k_n^2} k_n \cdot \mathcal A_n)-\frac{1}{2z}(\varepsilon_{1\nu}\partial_{k_1^{\mu}}  \mathcal{A}_n^{\nu} - \frac{\varepsilon_1^{\mu}}{k_1^2} k_1 \cdot \mathcal A_n),
}
in the second line, the derivative is expanded when acting on the helicity, to highlight that the terms exhibiting a $1/k_h^2$ pole correspond exactly to the OPE terms in the soft limit~\cite{Chowdhury:2024wwe}.
 We can also easily solve the constraint at any order up to an antisymmetric tensor as before, which gives:
\eqs{
\mathcal{A}_{n+1}=
&-\sum_{\ell =0 }^\infty\left( (z \varepsilon_q \cdot k_n+z \varepsilon_q^{\mu} q^{\nu} S_{n,\mu \nu}) \frac{1}{\mathcal{D}^{d-1}_{\vec k_n}}- \frac{1}{2 z(\ell + 1)} \varepsilon_q \cdot \partial_{k_n} \right) \frac{1}{\ell !}(q \cdot \partial_{k_n})^{\ell} \mathcal{A}_n 
\\
&
+
\left(n \leftrightarrow 1\right)
+ \varepsilon_q^{\mu} q^{\nu_1} \dots q^{\nu_{\ell}}N_{\mu \nu_1 \dots \nu_{\ell}}
}
where we rewrite the $\Sigma$ in eq.~\eqref{class1_ym} as a spin operator $S_{h,\mu \nu}= \varepsilon^{\mu} _h\partial_{\varepsilon_h}^{\nu}-\varepsilon_h^{\nu} \partial_{\varepsilon_h}^{\mu}$, and we have verified this formula up to 5 point amplitudes. However, clearly beyond the leading order we can not fully determine the soft operator, so we will only translate the leading order soft theorem into correlators in momentum space. Thus, following the same procedure from eq. \eqref{mmamp_soft_photon_theorem} to \eqref{eq25} we have:
\eqs{
\langle J(\vec q)J(\vec k_1) \dots J(\vec k_n)\rangle = \frac{1}{2}(\varepsilon_q \cdot \partial_{k_n}- \varepsilon_q \cdot \partial_{k_1})\langle J(\vec k_1) \dots J(\vec k_n) \rangle + \mathcal{O}(q).
}
This is also verified up to the five-point Yang-Mills correlator up to local terms.
\section{Soft Graviton theorems}
The three point Graviton amplitude is simply given by,
\eqs{
\mathcal{M}_3(1,2,3)=z^2(\varepsilon_1 \cdot \varepsilon_2 \varepsilon_3 \cdot k_1+\varepsilon_2 \cdot \varepsilon_3 \varepsilon_1 \cdot k_2 + \varepsilon_3 \cdot \varepsilon_1 \varepsilon_2 \cdot k_3)^2.
\label{gr3pt}
}
As before, we use only the factorization as our diagrammatic input for diagram (a), and other contributions like the OPE pole will be absorbed into the remainder function $R$,
\eqs{
\mathcal{M}_{n+1}=& \sum_{h=1}^n\sum_{r=\pm} \mathcal{M}_3(\vec q,\vec k_h,-\vec k_I^r) \frac{1}{\mathcal{D}_{\vec k_I}^{d}} \mathcal{M}_n
(\vec k_1, \dots, \vec k_I^{-r}, \dots, \vec k_n)
+R(\vec{q},\vec{k}_1,\ldots,\vec{k}_n)\\
=&\sum_{h=1}^n\mathcal{M}^{\mu \nu}_3
(\vec q,\vec k_h,-\vec k_I) \frac{ \Pi_{\mu \nu, \rho \sigma} }{\mathcal{D}_{\vec k_I}^{d}}  \mathcal{M}^{\rho \sigma}_n(\vec k_1, \dots, \vec k_I, \dots, \vec k_n)+R(\vec{q},\vec{k}_1,\ldots,\vec{k}_n). 
\label{gr_classI}
}
Putting the explicit three-point amplitude in, we can rewrite the first term as:
\eqs{
z^2[ \varepsilon_q \cdot k_h \varepsilon_h^{\mu}+ \varepsilon_q^{\mu_1}  q^{\mu_2} (\Sigma_{\mu_1 \mu_2})_{ \mu_3 \mu_4} \varepsilon_h^{\mu_3} \eta^{\mu \mu_4}] \times [ \mu \to \nu ] \times \Pi_{\mu \nu, \rho \sigma} \frac{1}{\mathcal{D}_{k_I}^d}\mathcal{M}^{\rho \sigma}_n(\vec k_I).
\label{eq37}
}
Similar to the photon example, there is only one term from $\mathcal{K}_q^a$ that can contribute to the $\frac{1}{q^2}$ pole, which is subsequently converted into an EoM operator to cancel the $1/\mathcal{D}_{\vec k_I}^d$ propagator. 
In particular, we find that ($\mathcal{K}_q^\mu+\mathcal{K}_h^\mu$) act on \eqref{eq37}: 
\eqs{
-\frac{4s_q+d}{q^2} \varepsilon_q^\mu(\varepsilon_q \cdot k_h \varepsilon_h^{\nu_1}+ \varepsilon_q^{\mu_1}  q^{\mu_2} (\Sigma_{\mu_1 \mu_2})_{ \mu_3 \mu_4} \varepsilon_h^{\mu_3} \eta^{\nu_1 \mu_4}) \varepsilon_h^{\nu_2} \Pi_{\nu_1 \nu_2, \rho \sigma} (\vec{k}_I)\mathcal{M}^{\rho \sigma}_n(\vec{k}_I)+\ldots
\label{eq39class1}
}
where $\ldots$ contains terms without $\frac{1}{q^2}$ poles.
Then for the remainder function $R$
\eqs{
\mathcal{K}^{\mu}_q (\varepsilon_q^{\nu_1} \varepsilon_q^{\nu_2} \cdot R_{\nu_1 \nu_2})=&\varepsilon_q^{\mu} \left( -\frac{4s_q+d}{q^2} (q^{\nu_1} \varepsilon_q^{\nu_2}  R_{\nu_1 \nu_2}+q^{\nu_2} \varepsilon_q^{\nu_1}  R_{\nu_1 \nu_2} )   \right) +\ldots
\label{eq40R}
}
\subsection{Solving SCWI in soft expansion}
The soft expansion of eq.~\eqref{gr_classI} gives:
\eqs{
&(\varepsilon_q \cdot k_h \varepsilon_h^{\mu}+ \varepsilon_q^{\mu_1}  q^{\mu_2} (\Sigma_{\mu_1 \mu_2})_{ \mu_3 \mu_4} \varepsilon_h^{\mu_3} \eta^{\mu \mu_4}) \varepsilon_h^{\nu} \Pi_{\mu \nu, \rho \sigma}(\vec k_I) \mathcal{M}^{\rho \sigma}_n(\vec k_1, \dots, \vec k_I, \dots, \vec k_n) \\
=& \sum_{\ell=0}^{\infty} \frac{1}{\ell !}( \varepsilon_q \cdot k_h + \frac{1}{2} \varepsilon_q^{\mu_1}  q^{\mu_2} S_{\mu_1 \mu_2} ) (q \cdot \partial_{k_h})^{\ell} \mathcal{M}_n(\vec k_1, \dots ,\vec k_h, \dots ,\vec k_n),\\
&\varepsilon_q^{\nu_1} \varepsilon_q^{\nu_2} \cdot R_{\nu_1 \nu_2}(\vec{q},\vec{k}_1,\ldots,\vec{k}_n)\\
=&(\varepsilon_q^{\nu_1} \varepsilon_q^{\nu_2} \cdot R^{(0)}_{\nu_1 \nu_2}(\vec{k}_1,\ldots,\vec{k}_n))+(\varepsilon_q^{\nu_1} \varepsilon_q^{\nu_2} q^{\nu_3} \cdot R^{(1)}_{\nu_1 \nu_2 \nu_3} (\vec{k}_1,\ldots,\vec{k}_n)) + \mathcal{O}\left(q^{2}\right)
\label{eq41softexp}
}
After soft expansion, at $\mathcal{O}(q^{-2})$ we see that only one term in eq.~\eqref{eq39class1} contributes,
\eqs{
\sum_{h=1}^n \left(-\frac{4s_q+d}{q^2}  \kappa_h \varepsilon_q \cdot k_h \right)\mathcal{M}_n=0.
}
This is only possible if the couplings are universal, and by momentum conservation,
\eqs{
\kappa_i= \kappa.
}
At $\mathcal{O}(q^{-1})$, we can get the constraint for $R^{(0)}$,
\eqs{
\sum_{h=1}^n\left( \varepsilon_q \cdot k_h (q \cdot \partial_{k_h})   
+\frac{1}{2} \varepsilon_q^{\mu} q^{\nu} S_{\mu \nu} 
\right) \mathcal{M}_n + q^{\mu} \varepsilon_q^{\nu} R^{(0)}_{\mu \nu}+q^{\nu} \varepsilon_q^{\mu} R^{(0)}_{\mu \nu}=0.
}
We can symmetrize and anti-symmetrize the index to obtain two solutions,
\eqs{
R^{(0)\mu \nu}=-\frac{1}{4} \sum_{h=1}^n (k_h^{\mu} \partial_{k_h}^{\nu}+ k_h^{\nu} \partial_{k_h}^{\mu}) \mathcal{M}_n(\vec k_1, \dots ,\vec k_h, \dots ,\vec k_n),\\
\sum_{h=1}^n(k_h^{\mu} \partial_{k_h}^{\nu}- k_h^{\nu} \partial_{k_h}^{\mu}+S^{\mu \nu})  \mathcal{M}_n(\vec k_1, \dots ,\vec k_h, \dots ,\vec k_n)  =0.
\label{eq45}
}
The second equation is simply the Ward identity from the boundary rotation generator.\\
At  $\mathcal{O}(q^{0})$, we can still fully determined the function $R^{(1)}$,
\eqs{
\sum_{h=1}^n\left(\varepsilon_q \cdot k_h  \frac{1}{2}( q \cdot \partial_{k_h})^2 
+ \frac{1}{2} \varepsilon_q^{\mu}  q^{\nu} S_{\mu \nu} (q \cdot \partial_{k_h}) 
\right)\mathcal{M}_n+ \frac{1}{2} \varepsilon_q^{\mu} q^{\nu} q^{\nu_1} (R^{(1)}_{\mu \nu \nu_1} +R^{(1)}_{\nu \mu \nu_1})=0.
}
This implies
\eqs{
 R^{{(1)}\mu \nu \nu_1}= \sum_{h=1}^n\left[-\frac{1}{4} (k_h^{\mu} \partial_{k_h}^{\nu}\partial_{k_h}^{\nu_1}+ k_h^{\nu} \partial_{k_h}^{\mu}\partial_{k_h}^{\nu_1}-\partial_{k_h}^{\mu}\partial_{k_h}^{\nu}k_h^{\nu_1} ) \mathcal{M}_n
 +\frac{1}{4} (S^{\mu \nu_1}\partial_{k_h}^{\nu}+S^{\nu \nu_1}\partial_{k_h}^{\mu})\right].
\label{eq47}
}
Collecting the solutions from constraints, in summary, we get the leading and sub-leading soft graviton theorem
\eqs{
\mathcal{M}_{n+1}=&\sum_{h=1}^{n} \bigg ( (z\varepsilon_q \cdot k_h)^2 \frac{1}{\mathcal{D}^d_{\vec k_I}} \mathcal{M}_n - \frac{1}{2}  (\varepsilon_q \cdot k_h \varepsilon_q \cdot \partial_{k_h})  \mathcal{M}_n\\
& +  (z \varepsilon_q \cdot k_h)^2 \frac{1}{\mathcal{D}^d_{\vec k_I}} (q \cdot \partial_{k_h})\mathcal{M}_n + z^2 \varepsilon_q \cdot k_h \varepsilon_q^{\mu} q^{\nu} S_{\mu \nu} \frac{1}{\mathcal{D}^d_{\vec k_I}} \mathcal{M}_n\\
&-\frac{1}{4}(2\varepsilon_q \cdot k_h \varepsilon_q \cdot \partial_{k_h} q\cdot \partial_{k_h}-q \cdot k_h (\varepsilon_q \cdot \partial_{k_h})^2 + 2\varepsilon^{\mu}_q q^{\nu} S_{\mu \nu} \varepsilon_q \cdot \partial_{k_h}) \mathcal{M}_n \bigg )+ \mathcal{O}(q^2).
}
We have verified this formula for four and five-graviton amplitudes from \cite{Mei:2023jkb,Mei:2024abu,Mei:2024sqz} is exactly correct. Then we should integrate out the bulk-to-bulk propagator to get a more compact result. In the end with soft limits $\vec{q} \to 0$ in $d=3$ we found,
\eqs{
\boxed{\lim_{\vec{q}\to 0}\langle T(\vec q) T(\vec k_1) \dots T(\vec k_n)\rangle =S^{(0)} \langle T(\vec k_1) \dots T(\vec k_n)\rangle + S^{(1)}\langle T(\vec k_1) \dots T(\vec k_n)\rangle + \mathcal{O}(q^2 )},
}
with the soft operator
\begin{equation}
 \begin{aligned}
   S^{(0)}:=&-\frac{1}{2}\sum_{a=1}^n \varepsilon_{\mu \nu} k_a^{\mu} \partial_{k_a}^{\nu},\\
    S^{(1)}:=&\frac{1}{4}\sum_{a=1}^n  \varepsilon_{\mu \nu} q_{\rho} (k_a^{\rho} \partial_{k_a}^{\mu} \partial_{k_a}^{\nu}-2k_a^{\mu} \partial_{k_a}^{\nu} \partial_{k_a}^{\rho}-2 \partial_{k_a}^{\mu}S_a^{\nu \rho}),
 \end{aligned}  
\end{equation}
where $S^{\mu \nu}=\varepsilon^{\mu} \partial_{\varepsilon}^{\nu}-\varepsilon^{\nu} \partial_{\varepsilon}^{\mu}$. We have verified this formula using the four-point graviton correlator from \cite{Bonifacio:2022vwa,Armstrong:2023phb} up to local terms\footnote{We remind the reader that $\partial_k^{\mu}$ should also act on helicity, and $\partial_{\varepsilon}^{\mu}$ will just strip of the helicity.}. It is interesting to notice the following properties of the soft graviton operators,
\eqs{
&S^{(0)} |_{\varepsilon^{\mu \nu} \to \eta^{\mu \nu}}=\sum_{a=1}^n D_a,\\
&S^{(1)} |_{\varepsilon^{\mu \nu} \to \eta^{\mu \nu}}=\sum_{a=1}^n q_{\rho} K_a^{\rho},
}
with the first one being the dilatation operator and the second one being special conformal generators. It is unclear what this operation means, but we hope such a nice feature would have a symmetry interpretation in the future.

\section{Comments on higher dimension operators}
We present a simple example to illustrate how a higher-dimensional operator modifies the soft theorem.
This will indicate whether the result is sensitive to UV physics.
Due to the presence of the AdS radius, simple dimensional analysis suggests that the modification will be very different from the flat space case~\cite{Bianchi:2014gla,Elvang:2016qvq,DiVecchia:2016amo}.
\\
Consider the following operator $\phi R^2$, whose three-point amplitude can be easily obtained by solving the special conformal Ward identity,
\eqs{
\mathcal{M}_3^{\phi R^2}(1^T,2^T,3^{\phi})=
&z^4
   (\varepsilon_1\cdot k_2)^2 (\varepsilon_2\cdot k_1)^2-4 \left(s_1+s_2\right) \left(d-2
   \left(s_1+s_2\right)\right) (\varepsilon_1\cdot \varepsilon_2)^2\\
&-2 z^2 \left(d+2-4 (s_1+ s_2)\right) \varepsilon_1\cdot \varepsilon_2 \varepsilon_1\cdot
   k_2 \varepsilon_2\cdot k_1,
}
which matches with \cite{Bzowski:2013sza}. Notice that the last two terms can be interpreted as curvature corrections to the flat-space S-matrix, and they contribute to the leading soft limit. Following the same steps as in the photon example, we find that at leading order, they modify the soft graviton theorem as follows in $d=3$,
\eqs{
\langle T(\vec q) T(\vec k_1) \dots T(\vec k_n)\rangle ^{\phi R^2}=-\frac{1}{2}\sum_{a=1}^n(\varepsilon_q \cdot \varepsilon_a)^2 \langle  T(\vec k_1)\dots\phi(\vec k_a) \dots T(\vec k_n)\rangle +\mathcal{O}(q ) .
}
\section{Outlook}
In this paper, we use part of the solution to the conformal Ward identities to derive soft theorems for photons, gluons, and gravitons in (A)dS space. In Minkowski space, the soft theorems have many new interpretations of symmetries. Thus, a natural direction is to ask: what are the corresponding Ward identities for these soft operators in (A)dS? In particular, our derivation is entirely from the bulk perspective. It would be very interesting to understand these results purely from the boundary point of view.

\begin{center}
\textbf{Acknowledgements}
\end{center}

We thank Mattia Arundine, Daniel Baumann, Calvin Chen, Harry Goodhew, Yu-Tin Huang, Callum Jones, Gordon Lee, Prahar Mitra, and Marko Simonović for many illuminating discussions. JM is supported by the European Union (ERC, UNIVERSE PLUS, 101118787). Views and opinions expressed are however those of the authors only and do not necessarily reflect those of the European Union or the European Research Council Executive Agency. Neither the European Union nor the granting authority can be held responsible for them.  YM is supported by a Edinburgh Global Research Scholarship.

\appendix

\section{Details on solving special conformal Ward identity for photon} \label{appendix:A}
We first consider $\mathcal{K}_q^{\mu}$ acting on the diagram (a),
\eqs{
&\mathcal{K}_q^{\mu} \left(  z \varepsilon_q \cdot k_h  G(\vec{k}_I,z,z')\mathcal{A}_n(z',\vec{k}_I, \dots) \right)\\
=&\varepsilon_q^{\mu} e_h G(\vec{k}_I,z,z') \left(-\frac{4s_q+d-2}{q^2} z q \cdot k_h \mathcal{A}_n(z',\vec{k}_I, \dots) +2z k_h \cdot \partial_q \mathcal{A}_n(z',\vec{k}_I, \dots)  \right)\\
&+\varepsilon_q^{\mu} e_h 4 z(q \cdot k_h+k_h^2) \frac{\partial G(\vec{k}_I,z,z')}{\partial k_{I}^2} \mathcal{A}_n(z',\vec{k}_I, \dots).
}
It's clear that only terms from the operator can contribute above, similarly for $\mathcal{K}_h^{\mu}$,
\eqs{
&\mathcal{K}^{\mu}_h \left(  z \varepsilon_s \cdot k_h G(\vec{k}_I,z,z') \mathcal{A}_n(z',\vec{k}_I, \dots) \right)\\
=&\varepsilon_q^{\mu} e_h G(\vec{k}_I,z,z') \left( (4s_h-d )z\mathcal{A}_n(z',\vec{k}_I, \dots) -2 z k_h \cdot \partial_q\mathcal{A}_n(z',\vec{k}_I, \dots)                \right)\\
&-\varepsilon_q^{\mu} e_h 4 z(k_s \cdot k_h+k_h^2) \frac{\partial G(\vec{k}_I,z,z')}{\partial k_{I}^2} \mathcal{A}_n(z',\vec{k}_I, \dots),  
}
then add them up together,
\eqs{
&(\mathcal{K}_q^{\mu}+\mathcal{K}^{\mu}_h ) \left(  z \varepsilon_s \cdot k_h G(\vec{k}_I,z,z') \mathcal{A}_n(z',\vec{k}_I, \dots) \right)\\
=&-\frac{4s_q+d-2}{q^2}\varepsilon_q^{\mu} e_h G(\vec{k}_I,z,z') \left(  z q \cdot k_h \mathcal{A}_n-\frac{q^2}{4s_q+d-2} (4s_h-d )z\mathcal{A}_n                \right)
\\
=&-\frac{4 s_q+d-2}{2 zq^2}\Big(z^2 k_I^2+2 (s_q+s_h) \big(d-2 (s_q+s_h+1)\big)\Big)  G(\vec{k}_I,z,z')  \varepsilon_q^\mu e_h \mathcal{A}_n.
\\
}
As explained in Appendix~H of~\cite{Mei:2024sqz}, this corresponds exactly to the identity operator. The procedure simply requires integration by parts while discarding the boundary term---which we identify as the inhomogeneous term appearing on the right-hand side of the special conformal Ward identities for spinning correlators.
\eqs{
\left(\mathcal{K}_q^\mu+\mathcal{K}_h^\mu\right) \left(  z \varepsilon_s \cdot k_h G(\vec{k}_I,z,z') \mathcal{A}_n \right)=-\frac{4 s_q+d-2}{2 zq^2}\mathcal{D}_{k_I}^{\Delta}  G(\vec{k}_I,z,z')  \varepsilon_q^\mu e_h \mathcal{A}_n,
}
then by definition of the bulk-to-bulk $\mathcal{D}_k^{\Delta} G(z, z')=z^{d+1} \delta(z-z') $. We get the delta function and then integrate out $z'$ gives eq~\eqref{eq_ideneity_op_photon}.

\bibliography{references.bib} 
\bibliographystyle{JHEP}

\end{document}